\documentclass[twocolumn,tighten,trackchanges]{aastex631}

\shorttitle{}
\shortauthors{Selvi et al.}

\graphicspath{{./}{figures/}}

\usepackage{longtable}

\usepackage{array,booktabs}
\usepackage[symbols,nogroupskip,nonumberlist]{glossaries-extra}

\usepackage[normalem]{ulem}
\usepackage{soul}
\usepackage{array}

\usepackage{mathrsfs}

\usepackage{xcolor}

\glsdisablehyper
\glsxtrnewsymbol[description={Speed of light}]{spol}{\ensuremath{c}}
\glsxtrnewsymbol[description={Time}]{time}{\ensuremath{t}}
\glsxtrnewsymbol[description={Number density}]{ndens}{\ensuremath{n}}
\glsxtrnewsymbol[description={Number}]{num}{\ensuremath{N}}
\glsxtrnewsymbol[description={Volume}]{vol}{\ensuremath{V}}
\glsxtrnewsymbol[description={Mass}]{mass}{\ensuremath{M}}
\glsxtrnewsymbol[description={Mass density}]{mdens}{\ensuremath{\rho}}
\glsxtrnewsymbol[description={Generalized coordinates}]{gcoords}{\ensuremath{\textbf{q}}}
\glsxtrnewsymbol[description={Generalized coordinate}]{gcoord}{\ensuremath{q}}
\glsxtrnewsymbol[description={Magnetic field energy density}]{magnenrgdens}{\ensuremath{\mathcal{E}_{\mathrm{B}}  }}
\glsxtrnewsymbol[description={Magnetic field components}]{Bcomp}{\ensuremath{B}}

\glsxtrnewsymbol[description={Magnetic field 3-vector}]{B}{\ensuremath{\textbf{B}}}
\glsxtrnewsymbol[description={Electric field energy density}]{elecenrgdens}{\ensuremath{\mathcal{E}_{\mathrm{E}}}}
\glsxtrnewsymbol[description={Electric field components}]{Ecomp}{\ensuremath{E}}
\glsxtrnewsymbol[description={Electric field 3-vector}]{E}{\ensuremath{\textbf{E}}}
\glsxtrnewsymbol[description={Electromagnetic field energy density}]{elecmagnenrgdens}{\ensuremath{\mathcal{E}_{\mathrm{EM}}}}
\glsxtrnewsymbol[description={Lorentz factor}]{lfacp}{\ensuremath{\gamma}}
\glsxtrnewsymbol[description={Lorentz factor}]{lfacbulk}{\ensuremath{\Gamma}}
\glsxtrnewsymbol[description={Particle mass}]{pmass}{\ensuremath{m_{p}}}
\glsxtrnewsymbol[description={Particle energy}]{prtenrgdens}{\ensuremath{E_{\mathrm{p}}}}
\glsxtrnewsymbol[description={Particle rest-mass energy}]{rmenrgdens}{\ensuremath{E_{\mathrm{rm}}}}
\glsxtrnewsymbol[description={Particle kinetic energy}]{kinenrgdens}{\ensuremath{E_{\mathrm{k}}}}
\glsxtrnewsymbol[description={3-velocity components}]{3velcomp}{\ensuremath{v}}
\glsxtrnewsymbol[description={3-velocity}]{3vel}{\ensuremath{\textbf{v}}}
\glsxtrnewsymbol[description={4-velocity components}]{4velcomp}{\ensuremath{u}}
\glsxtrnewsymbol[description={4-momentum components}]{4momcomp}{\ensuremath{p}}
\glsxtrnewsymbol[description={4-velocity}]{4vel}{\ensuremath{\textbf{u}}}
\glsxtrnewsymbol[description={4-momentum}]{4mom}{\ensuremath{\textbf{p}}}
\glsxtrnewsymbol[description={Energy-momentum tensor}]{enrgmomten}{\ensuremath{\textbf{T}}}
\glsxtrnewsymbol[description={Energy-momentum tensor components}]{enrgmomtencomp}{\ensuremath{T}}
\glsxtrnewsymbol[description={Energy-momentum tensor density components}]{enrgmomtendenscomp}{\ensuremath{\mathfrak{T}}}
\glsxtrnewsymbol[description={4-position}]{4pos}{\ensuremath{\mathbf{x}}}
\glsxtrnewsymbol[description={4-position components}]{4poscomp}{\ensuremath{x}}
\glsxtrnewsymbol[description={Metric tensor}]{metr}{\ensuremath{\textbf{g}}}
\glsxtrnewsymbol[description={Metric tensor components}]{metrcomp}{\ensuremath{g}}
\glsxtrnewsymbol[description={Affine parameter}]{affpar}{\ensuremath{s}}
\glsxtrnewsymbol[description={System distribution function}]{distrfunc}{\ensuremath{f}}
\glsxtrnewsymbol[description={Metric tensor}]{metrten}{\ensuremath{\textbf{g}}}
\glsxtrnewsymbol[description={Metric tensor components}]{metrtencomp}{\ensuremath{\tilde{g}}}
\glsxtrnewsymbol[description={Metric tensor determinant}]{metrtendet}{\ensuremath{g}}
\glsxtrnewsymbol[description={Minkowksi metric tensor components}]{minkmetrtencomp}{\ensuremath{\eta}}
\glsxtrnewsymbol[description={Cartesian x coordinate}]{cartx}{\ensuremath{x}}
\glsxtrnewsymbol[description={Cartesian y coordinate}]{carty}{\ensuremath{y}}
\glsxtrnewsymbol[description={Cartesian z coordinate}]{cartz}{\ensuremath{z}}
\glsxtrnewsymbol[description={Spherical radial coordinate}]{sphradius}{\ensuremath{r}}
\glsxtrnewsymbol[description={Spherical polar coordinate}]{sphpolar}{\ensuremath{\theta}}
\glsxtrnewsymbol[description={Spherical azimuthal coordinate}]{sphazi}{\ensuremath{\phi}}
\glsxtrnewsymbol[description={Schwarzschild radius}]{schwradius}{\ensuremath{R_{s}}}
\glsxtrnewsymbol[description={Graviational constant}]{gravconst}{\ensuremath{G}}
\glsxtrnewsymbol[description={Black hole spin parameter}]{spinpar}{\ensuremath{a}}
\glsxtrnewsymbol[description={Magnetic vector potential}]{magnvecpot}{\ensuremath{\textbf{A}}}
\glsxtrnewsymbol[description={Magnetic vector potential components}]{magnvecpotcomp}{\ensuremath{A}}
\glsxtrnewsymbol[description={Current}]{curr}{\ensuremath{I}}
\glsxtrnewsymbol[description={Vacuum mangetic permeability}]{vacmagnperm}{\ensuremath{\mu_{0}}}
\glsxtrnewsymbol[description={Magnetic dipole moment}]{magndipmom}{\ensuremath{ \textbf{m} }}

\glsxtrnewsymbol[description={Electromagnetic 4-potential}]{4pot}{\ensuremath{\textbf{A}}} 
\glsxtrnewsymbol[description={Electromagnetic 4-potential components}]{4potcomp}{\ensuremath{A}} 
\glsxtrnewsymbol[description={Electric scalar potential}]{elecpot}{\ensuremath{ \tilde{\phi}  }} 
\glsxtrnewsymbol[description={Electromagnetic field tensor}]{emtens}{\ensuremath{\textbf{F}}} 
\glsxtrnewsymbol[description={Electromagnetic field tensor components}]{emtenscomp}{\ensuremath{F}} 
\glsxtrnewsymbol[description={Electrical resistivity}]{elres}{\ensuremath{\eta}} 
\glsxtrnewsymbol[description={Magnetic diffusivity}]{mdiff}{\ensuremath{\tilde{\eta}}} 
\glsxtrnewsymbol[description={Electrical conductance}]{elcond}{\ensuremath{\sigma_{e}}} 
\glsxtrnewsymbol[description={Modified Kerr-Schild radius stretching parameter}]{mksr}{\ensuremath{R_{0,MKS}}} 
\glsxtrnewsymbol[description={Modified Kerr-Schild polar stretching parameter}]{mksh}{\ensuremath{h_{0,MKS}}} 
\glsxtrnewsymbol[description={Plasma magnetization}]{plmagn}{\ensuremath{\sigma}} 
\glsxtrnewsymbol[description={Electromagnetic vector potential}]{vecpot}{\ensuremath{\textrm{A} }} 
\glsxtrnewsymbol[description={Electromagnetic vector potential components}]{vecpotcomp}{\ensuremath{A}} 
\glsxtrnewsymbol[description={Magnetic moment}]{magnmom}{\ensuremath{\mu}} 
\glsxtrnewsymbol[description={Pressure}]{pres}{\ensuremath{p}} 
\glsxtrnewsymbol[description={Temperature}]{temp}{\ensuremath{T}} 

\glsxtrnewsymbol[description={Magnetopheric tilt angle}]{tiltangle}{\ensuremath{\chi}} 
\glsxtrnewsymbol[description={Length}]{len}{\ensuremath{\mathcal{L}}} 
\glsxtrnewsymbol[description={Resolution}]{res}{\ensuremath{\mathcal{R}}} 
\glsxtrnewsymbol[description={Plasma beta}]{plbeta}{\ensuremath{\beta}} 

\glsxtrnewsymbol[description={Fluid pressure}]{flpr}{\ensuremath{p_{f}}} 
\glsxtrnewsymbol[description={Polytropic index}]{polyidx}{\ensuremath{\hat{\gamma}}} 
\glsxtrnewsymbol[description={Entropy related quantity}]{entralt}{\ensuremath{S}} 
\glsxtrnewsymbol[description={Entropy related quantity}]{boltzcon}{\ensuremath{k_{\mathrm{B}}}} 
\glsxtrnewsymbol[description={Dimensionless temperature/thermal spread}]{dimtemp}{\ensuremath{\Theta}} 
\glsxtrnewsymbol[description={Magnetic pressure}]{magnpr}{\ensuremath{p_{\mathrm{m}}}} 
\glsxtrnewsymbol[description={Vector potential constant}]{vecpotcon}{\ensuremath{g}} 
\glsxtrnewsymbol[description={Magnetic inclination angle}]{magninclangle}{\ensuremath{\chi}} 
\glsxtrnewsymbol[description={Relativistic enthaply density}]{relenthdens}{\ensuremath{w}} 
\glsxtrnewsymbol[description={Gravitational radius}]{gravradius}{\ensuremath{r_{\textrm{g}}  }} 
\glsxtrnewsymbol[description={Angular velocity}]{angfreq}{\ensuremath{ \Omega}} 
\glsxtrnewsymbol[description={Volume 3-current density}]{3currdenscomp}{\ensuremath{J}}
\glsxtrnewsymbol[description={Magnetic flux}]{magnflux}{\ensuremath{ \Phi    }}
\glsxtrnewsymbol[description={Volume 3-current density}]{3currdens}{\ensuremath{\textbf{J}}}
\glsxtrnewsymbol[description={Alfv{\'e}n speed}]{alfvensp}{\ensuremath{  v_{\textrm{A}}   }}
\glsxtrnewsymbol[description={Drift velocity}]{driftvel}{\ensuremath{  v_{\textrm{d}}   }}
\glsxtrnewsymbol[description={Spatial metric determinant}]{spmetricdet}{\ensuremath{  \gamma  }}
\glsxtrnewsymbol[description={Radial MKS coordinate}]{mksradius}{\ensuremath{  s  }}
\glsxtrnewsymbol[description={Lapse function}]{lapse}{\ensuremath{  \alpha  }}
\glsxtrnewsymbol[description={Power}]{power}{\ensuremath{  P  }}
\glsxtrnewsymbol[description={Energy flux}]{enrgyflux}{\ensuremath{  \dot{E} }}
\glsxtrnewsymbol[description={Event horizon radius}]{ehradius}{\ensuremath{   r_{\textrm{H} } }} 
\glsxtrnewsymbol[description={Magnetic flux}]{magnfluxabs}{\ensuremath{ \Phi_{|\textrm{B}|}   }}
\glsxtrnewsymbol[description={Angular momentum}]{angmom}{\ensuremath{\textbf{L}}} 
\glsxtrnewsymbol[description={Comoving magnetic field components}]{bcomp}{\ensuremath{b}}
\glsxtrnewsymbol[description={Comoving magnetic field 3-vector}]{b}{\ensuremath{\textbf{b}}}
\glsxtrnewsymbol[description={Period}]{period}{\ensuremath{P}}
\glsxtrnewsymbol[description={Elementary charge}]{elch}{\ensuremath{e}}
\glsxtrnewsymbol[description={Mulitplicity}]{mplcty}{\ensuremath{\lambda}}
\glsxtrnewsymbol[description={Force}]{force}{\ensuremath{F}}
\glsxtrnewsymbol[description={Reconnection rate}]{recrate}{\ensuremath{R}}
\glsxtrnewsymbol[description={Energy density}]{enrgdens}{\ensuremath{U}}
\glsxtrnewsymbol[description={Thomson cross section}]{csthomson}{\ensuremath{\sigma_{\textrm{T}}}}
\glsxtrnewsymbol[description={Length}]{length}{\ensuremath{L}}
\glsxtrnewsymbol[description={Energy}]{enrg}{\ensuremath{\mathcal{E}}}
\glsxtrnewsymbol[description={Relativistic specific enthaply}]{relspecenth}{\ensuremath{h}} 
\glsxtrnewsymbol[description={Magnetic field components co-moving}]{Bcocomp}{\ensuremath{b}}
\glsxtrnewsymbol[description={delzanna efield}]{cEcomp}{\ensuremath{ \mathcal{E} }} 
\glsxtrnewsymbol[description={Transport velocity components}]{trvelcomp}{\ensuremath{\mathcal{V}}}
\glsxtrnewsymbol[description={Shift vector}]{shiftv}{\ensuremath{ \tilde{\beta}  }}
\glsxtrnewsymbol[description={Surface area}]{area}{\ensuremath{ S }}
\glsxtrnewsymbol[description={Surface area fraction}]{areafrac}{\ensuremath{ \tilde{S} }}

\glsxtrnewsymbol[description={Cumulative surface coordinate}]{cumsurfcoord}{\ensuremath{ \varsigma }}
\glsxtrnewsymbol[description={Spatial metric components}]{spmetriccomp}{\ensuremath{  \gamma  }}

\glsxtrnewsymbol[description={Factor}]{factor}{\ensuremath{  f  }}
\glsxtrnewsymbol[description={Solar mass}]{solmass}{\ensuremath{M_{\odot}}}
\glsxtrnewsymbol[description={Radius}]{radius}{\ensuremath{R}}
\glsxtrnewsymbol[description={Timescale}]{tscale}{\ensuremath{\tau}}
\glsxtrnewsymbol[description={Multipole moment}]{mpolemoment}{\ensuremath{\mu}}

\begin{document}

 
\title{On the universality of the split monopole black hole  magnetosphere}

\author[0000-0001-9508-1234]{S. Selvi}
\affiliation{Department of Astronomy, Columbia University, 550 West 120th street, New York, NY, 10027, USA}
\affiliation{Anton Pannekoek Institute, Science Park 904, 1098 XH, Amsterdam, The Netherlands}

\author[0000-0002-4584-2557]{O. Porth}
\affiliation{Anton Pannekoek Institute, Science Park 904, 1098 XH, Amsterdam, The Netherlands}

\author[0000-0002-7301-3908]{B. Ripperda}
\affiliation{Canadian Institute for Theoretical Astrophysics, 60 St. George Street, Toronto, ON M5S 3H8, Canada}
\affiliation{Department of Physics, University of Toronto, 60 St. George Street, Toronto, ON M5S 1A7, Canada}
\affiliation{David A. Dunlap Department of Astronomy, University of Toronto, 50 St. George Street, Toronto, ON M5S 3H4, Canada}
\affiliation{Perimeter Institute for Theoretical Physics, 31 Caroline St. North, Waterloo, ON N2L 2Y5, Canada }

\author[0000-0002-1227-2754]{L. Sironi}
\affiliation{Department of Astronomy, Columbia University, 550 West 120th street, New York, NY, 10027, USA}
\affiliation{Center for Computational Astrophysics, Flatiron Institute, 162 5th avenue, New York, NY, 10010, USA}

\begin{abstract}
Black holes can acquire magnetic flux from their magnetized progenitor or via prolonged accretion.  
We study the evolution of black hole magnetospheres by means of axisymmetric general relativistic magnetohydrodynamic simulations. 
We show that all simulated initial magnetic field geometries of varying complexity ultimately evolve into a split monopole magnetosphere.
The magnetospheric evolution consists of two phases. 
In the first phase, the magnetosphere evolves toward pressure equilibrium accompanied by a large magnetic flux decrease on the event horizon on a fast Alfv\'enic timescale of $\sim 60$ light-crossing times of the gravitational radius.   
The second phase proceeds in a pressure balance in which the magnetic flux decays and current sheets shift in polar angle over the event horizon on slower resistive timescales. 
We present an analytic model for the second phase.
Furthermore, we show that in a split monopole magnetosphere the magnetic flux on the event horizon decays exponentially with a timescale that depends on the black hole spin, where higher spin results in slower decay.  
Our results can have an implications for the timescales of reconnection-powered flares and for multimessenger counterparts to gravitational wave events.
\end{abstract}

\keywords{
\href{http://astrothesaurus.org/uat/739}{High energy astrophysics (739)}; 
\href{http://astrothesaurus.org/uat/1261}{Plasma astrophysics (1261)}; 
\href{http://astrothesaurus.org/uat/288}{Compact objects (288)}; 
\href{http://astrothesaurus.org/uat/994}{Magnetic fields (994)}; 
\href{http://astrothesaurus.org/uat/1393}{Relativity (1393)}; 
\href{http://astrothesaurus.org/uat/1964}{Magnetohydrodynamics (1964)} 
}

\section{Introduction}
Black hole (BH) magnetospheres are associated with many spectacular high-energy phenomena such as
BH jets, BH--neutron star post-merger emission, and BH magnetospheric flares. 
For actively accreting BHs, the inflow of plasma drives electric currents that sustain the magnetic flux threading the event horizon.
Non-accreting BHs can acquire magnetic flux from their magnetized progenitors \citep{Falcke2014} or by merging with a magnetized star (e.g., a neutron star (NS) \citep{East2021}). 
Furthermore, a BH in the magnetically arrested disk (MAD) state acquires magnetic flux on its event horizon from accreting plasma. 
When the magnetic pressure gradient of the poloidal open field lines becomes stronger than the gravitational pull on the accreting plasma, the accretion disk is pushed away and forms a transient magnetosphere while momentarily suppressing accretion \citep{Porth2021,Ripperda2022}. 

In vacuum, on light-crossing timescales magnetic flux on a BH event horizon will inevitably decay \citep{Price1972} according to the BH no-hair theorem \citep{Misner1973}. 
However, in nature, BHs that acquire magnetic flux will very likely be surrounded by a highly conducting plasma, thereby changing the magnetospheric dynamics compared to the vacuum case. 
Poloidal magnetic field loops with their footpoints penetrating the event horizon in a stationary axisymmetric magnetosphere can only exist if supported by non-force-free currents \citep{MacDonald1982,Gralla2014}. 
As a result, a force-free BH magnetosphere with a dipole magnetic field geometry will evolve into a split monopole geometry \citep{Komissarov2004,Lyutikov2011,Bransgrove2021} in which the current sheet extends all the way to the event horizon. 
An extremely high, but finite, plasma conductivity inevitably leads magnetic reconnection to dissipate magnetic field energy in the equatorial current sheet of the split monopole magnetosphere, leading to event horizon magnetic flux decay \citep{Lyutikov2011,Bransgrove2021,Selvi2024}, thereby extending the BH no-hair theorem from vacuum to the presence of a plasma.
 
Previous studies investigated the magnetospheric evolution when the BH magnetic and rotational axes were aligned for an initially dipolar magnetic field \citep{Bransgrove2021} or misaligned for an initially split monopolar magnetic field \citep{Selvi2024}. 
However, the magnetic field acquired by a BH likely has a more complex structure. 
The magnetic field inherited by the BH from a magnetized progenitor or a merging star typically does not have a purely dipolar geometry centered on the BH, but contains multipolar components and can be off-centered from the BH's center.
Similarly, during accretion, the magnetic field is not purely dipolar but shaped by the complex structure of the surrounding accretion flow.

In this work, we study the dynamics of force-free BH magnetospheres endowed with a magnetic field structure of varying complexity. 
We present an analytic model for the magnetospheric dynamics during the evolutionary phase in which pressure balance is established and evolution is controlled by magnetic reconnection.
We conclude that any (non-equilibrium) magnetic field configuration ultimately evolves into a split monopole magnetosphere.
Furthermore, we study the magnetic flux decay on the event horizon during a split monopole geometry and its dependence on the BH spin showing that flux decay slows down for higher BH spin.

\section{Numerical methods and setup} 
\label{sec:numericalmethods} 
We adopt a general relativistic ideal magnetohydrodynamic (GRMHD) description of the magnetospheric plasma and employ a fixed Kerr spacetime geometry for a range of dimensionless BH spins $\gls{spinpar} \in \{0,0.5,0.8,0.9,0.99 \}$.
We denote the BH mass as $\gls{mass}$, the gravitational constant as $\gls{gravconst}$, the speed of light as $\gls{spol}$, the gravitational radius as $\gls{gravradius} = \gls{gravconst} \gls{mass} / \gls{spol}^{2}$, and the event horizon radius $\gls{sphradius}_{\textrm{H}} = (1 + \sqrt{1 - \gls{spinpar}^{2}} ) \gls{gravradius}$.
The simulations are performed using the BlackHoleAccretionCode (\texttt{BHAC}) \citep{Porth2017,Olivares2019}.
The GRMHD equations are solved numerically in axisymmetric spherical, event horizon penetrating logarithmic Kerr-Schild coordinates $(\gls{time}, \gls{mksradius},\gls{sphpolar},\gls{sphazi})$ \citep{McKinney2004,Porth2017}, where $\gls{time}$ is time,
$\gls{mksradius}=\textrm{ln}(\gls{sphradius})$; $\gls{sphradius}$, $\gls{sphpolar}$, $\gls{sphazi}$ are the spatial spherical Kerr-Schild coordinates.
The BH spin axis is aligned with the $\gls{cartz}$-axis (i.e., along $\gls{sphpolar} = 0$). 
To vary the degree of complexity, we initialize the Eulerian magnetic field $\gls{Bcomp}^{i}$ with  $i \in \{ \gls{mksradius},\gls{sphpolar},\gls{sphazi} \}$ by magnetic vector potentials for a dipole, an offset dipole, a quadrupole, a quadrudipole (i.e., combination of dipole and quadrupole), and an octuquadrupole (i.e., combination of quadrupole and octupole). 
The strength of a multipole of order $l$ is determined by its multipole moment $\gls{mpolemoment}_{l}$ (e.g., $\gls{mpolemoment}_{1}$ the dipole moment, $\gls{mpolemoment}_{2}$ the quadrupole moment, $\gls{mpolemoment}_{3}$ the octupole moment). 
The relative strength of the multipoles can be quantified by their polar magnetic field strength $\gls{Bcomp}^{(l)}_{\textrm{p}}$ at the event horizon, which are related as
$
\gls{Bcomp}^{(l)}_{\textrm{p}}
=
(\gls{mpolemoment}_{l}/\gls{mpolemoment}_{k}) \gls{ehradius}^{k-l}
\gls{Bcomp}^{(k)}_{\textrm{p}}
$, where $l$ and $k$ are the order of the multipoles and $\gls{mpolemoment}_{l} / \gls{mpolemoment}_{k}$ has units of $\gls{gravradius}^{l-k}$.
For example, for a non-spinning BH having a quadrudipole with  $\gls{mpolemoment}_{1} / \gls{mpolemoment}_{2} = (1/30) \gls{gravradius}^{-1} $ the polar field strengths on the event horizon of the quadrupole and dipole part relate as $\gls{Bcomp}^{(1)}_{\textrm{p}} =\gls{Bcomp}^{(2)}_{\textrm{p}} /15$. 
We adopt the equation of state of a fully relativistic ideal gas with adiabatic index $\gls{polyidx} = 4/3$.
We initialize a pressure $\gls{pres}  = \gls{dimtemp} \gls{mdens}  \gls{spol}^{2}$, where $\gls{mdens}$ is the plasma rest-mass density and $\gls{dimtemp} = 0.05$ (i.e., a cold plasma) is the dimensionless temperature. 
See Appendix \ref{app:setup} for more details on the initialization and numerical floors of the numerical setup.

The computational domain covers $ 1.01  \leq \gls{sphradius} /  \gls{gravradius} \leq 400 $ and $0 \leq \gls{sphpolar}  \leq \pi $ and we run all simulations until time $\gls{time} = 200 \ \gls{gravradius} / \gls{spol}$.
We apply a second order accurate ``Koren'' reconstruction scheme as well as second order time-integration \citep{Porth2017,Olivares2019}. 
We use a uniform computational grid with a base resolution of $512 \times 256$ cells in $\gls{mksradius} \times \gls{sphpolar}$ and increase resolution by consecutively doubling the number of cells in each coordinate direction.
Our convergence study (see Appendix \ref{app:numconv}) up to a resolution of $16384 \times 8192$ cells in $\gls{mksradius}\times\gls{sphpolar}$ indicates converged results upward from $2048 \times 1024$ cells in $\gls{mksradius}\times\gls{sphpolar}$ and consequentially all the results reported here use at least this resolution. 
We report results from four simulations of a non-spinning BH ($\gls{spinpar}=0$) varying the initial magnetic field geometry using $4096 \times 2048$ cells in $\gls{mksradius}\times\gls{sphpolar}$.
Furthermore, in four more simulations with an initial dipole magnetic field geometry we vary the BH spin using $2048 \times 1024$ cells in $\gls{mksradius}\times\gls{sphpolar}$.
In our converged simulations reconnection occurs in the plasmoid-mediated regime such that the reconnection rate becomes asymptotically independent of resolution.
We employ cgs units.

\section{Magnetospheric evolution} 
\label{sec:magnetopshericevolution}
\begin{figure*}[ht!]
\centering
\resizebox{\textwidth}{!}{ 
\plotone{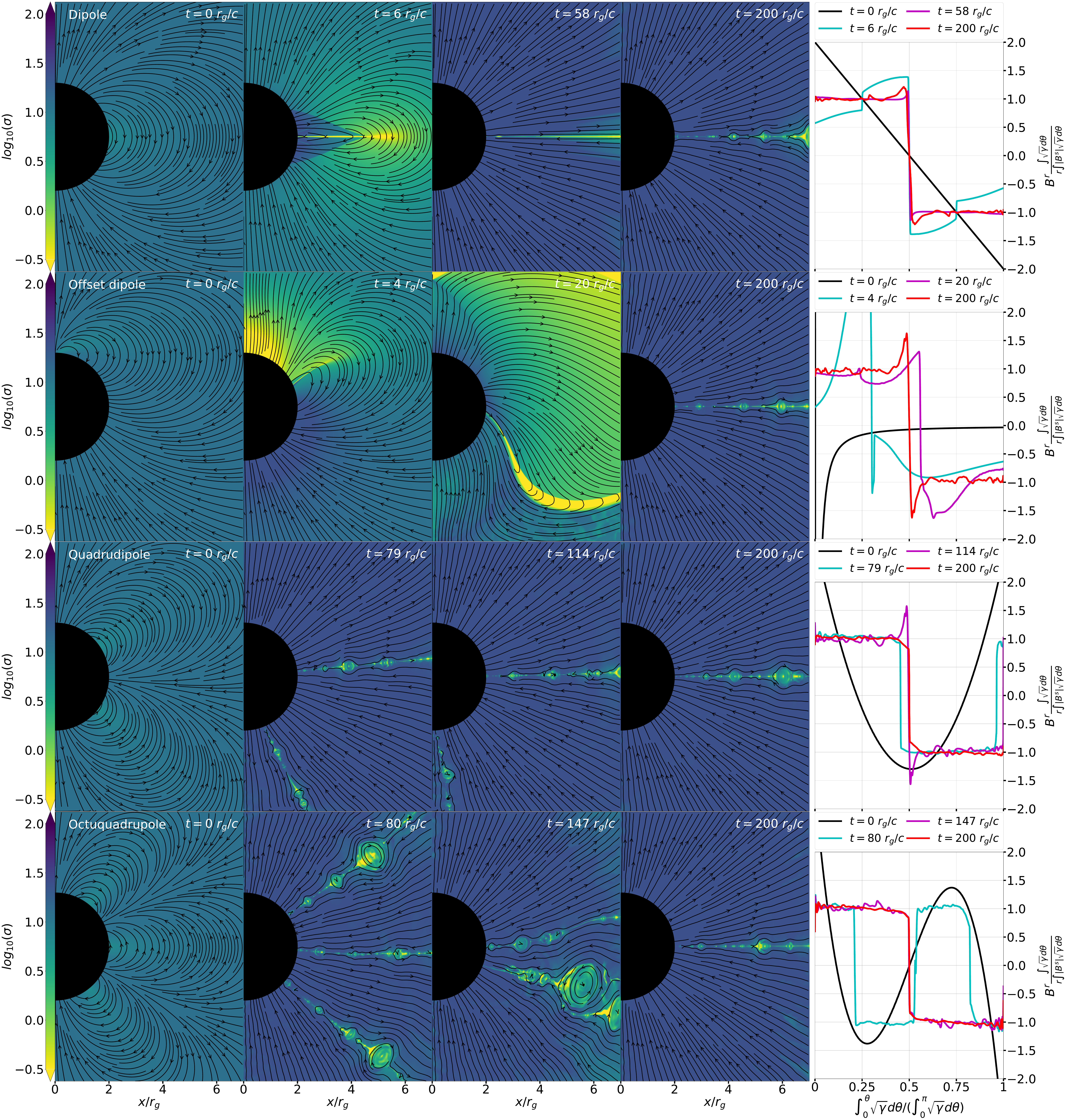}
}
\caption{   
Time sequences of the dipole, offset dipole, quadrudipole, and octuquadrupole initial magnetic field geometries in respectively the first to fourth row.
The four panels on the left for four individually selected times display the plasma magnetization $\gls{plmagn}$ in the 2D plane.
The BH shown by the black circle.
Magnetic field lines are represented by black arrowed lines.
}
\label{fig:panels}
\end{figure*}

In this section we consider a non-spinning BH with a dipole, offset dipole, quadrudipole, and octuquadrupole initial magnetic field geometry.
Figure \ref{fig:panels} displays for each case in each row a time sequence of the 2D structure of the plasma magnetization
$
\gls{plmagn}
=
\gls{bcomp}^{2} /( 4 \pi  \gls{mdens}
\gls{relspecenth})  
$ with 
$\gls{relspecenth}
=
( 1 + 
 (\gls{polyidx}/(\gls{polyidx} -1)) 
\gls{dimtemp}  )   \gls{spol}^{2} 
$ the relativistic specific enthalpy and
$\gls{bcomp}^{i}$ the co-moving magnetic field. 
The times are selected to include the initial state at $\gls{time}=0 \ \gls{gravradius} / \gls{spol}$ in the first column, intermediate states at illustrative times specific for each case in the second and third column, and the final state at $\gls{time} = 200 \ \gls{gravradius} / \gls{spol}$ in the fourth column. 
The right column show for the four selected times the normalized radial magnetic field distribution $\gls{Bcomp}^{\gls{sphradius}}  \int_{0}^{2 \pi} \int_{0}^{\pi} \sqrt{\gls{spmetricdet}} d\gls{sphpolar} d\gls{sphazi}  / (\gls{sphradius} \int_{0}^{2 \pi} \int_{0}^{\pi} |\gls{Bcomp}^{\gls{mksradius}}| \sqrt{\gls{spmetricdet}} d\gls{sphpolar}d\gls{sphazi})$ over the event horizon as a function of the normalized cumulative surface coordinate $\gls{cumsurfcoord} = \int_{0}^{\gls{sphpolar}} \sqrt{\gamma} d \gls{sphpolar} / (\int_{0}^{\pi} \sqrt{\gamma} d \gls{sphpolar})$ 
with 
$\gls{Bcomp}^{\gls{sphradius}} = \gls{Bcomp}^{\gls{mksradius}}  \gls{sphradius}$ the magnetic field in the radial Kerr-Schild coordinate 
and 
$\gls{spmetricdet}$ the determinant of the spatial part of the metric.
In the normalization of the radial magnetic field the numerator holds the total event horizon surface area 
$\gls{area}_{\textrm{tot}} = \int_{0}^{2 \pi} \int_{0}^{\pi} \sqrt{\gls{spmetricdet}} d\gls{sphpolar} d\gls{sphazi} $ 
and the denominator the instantaneous absolute magnetic field flux on the event horizon $\gls{magnflux}_{\textrm{tot}}(\gls{time}) = \int_{0}^{2 \pi} \int_{0}^{\pi} |\gls{Bcomp}^{\gls{mksradius}}(\gls{time})|\sqrt{\gls{spmetricdet}} d\gls{sphpolar}d\gls{sphazi}$. 
Normalizing the radial magnetic field in this way gives the advantage that in pressure equilibrium its magnitude is equal to unity, everywhere apart from current sheets.

\paragraph{Dipole}
First we consider the case of an initial dipole geometry (see in Figure \ref{fig:panels} the first row). 
Initially, poloidal magnetic field loops contract and fall into the BH
(see in Figure \ref{fig:panels} in the first row the first and second panels).  
Simultaneously, flux moves from the polar regions to the equatorial region, squeezing poloidal field loops near the equator (see in Figure \ref{fig:panels} in the first row the second and third panels). 
The remaining magnetic field is purely radial and develops a strong gradient in $\gls{sphpolar}$-direction at the equatorial plane, thereby forming a current sheet and settling into a split monopole geometry after $\sim 60 \ \gls{gravradius} / \gls{spol}$.

We characterize the topology of the magnetic field as the number of regions with ingoing and outgoing flux on the event horizon, irrespective of their specific properties (i.e., angle localization, occupied surface area, and bounding contour). 
For the initial dipole, the topology on the event horizon is such that there is one region of outgoing flux covering one hemisphere and one region of ingoing flux covering the other hemisphere. 
Therefore, the largest poloidal field loops extend over the entire polar angle.
Although during the evolution the dipole geometry changes to a split monopole geometry, the magnetic field topology on the event horizon remains the same.

The split monopolar equatorial current sheet becomes susceptible to the tearing instability leading to the formation of x-points. 
Herein, magnetic reconnection dissipates magnetic field energy to the fluid, thereby weakening the magnetic field strength causing magnetic flux decay.
Abundant x-point generation leads to the formation of plasmoids (i.e., two-dimensional cross sections of three-dimensional magnetic flux tubes) developing a hierarchical plasmoid chain in which the plasmoids advect, merge and grow (see in Figure \ref{fig:panels}) and frequently generating monster-sized plasmoids from the inner magnetospheric region.

\paragraph{Offset dipole}
The initial offset dipole is not centered on the BH like the dipole discussed before, but is off-centered from the BH's center and located at $(\gls{sphradius},\gls{sphpolar}) = (1.9 \gls{gravradius},0)$ just within the event horizon. 
The topology of the offset dipole is the same as that of the dipole as there is one region of each flux polarity. 
However, due to the offset the two regions are not symmetrically distributed over the event horizon, but the surface area of one is small and the other large. 
Similar to the centered dipole case, poloidal field loops fall into the BH and flux moves from the polar regions to the equatorial region squeezing magnetic field lines towards the equator and forcing field lines along the radial direction.
Due to the offset of the dipole breaking the symmetry of the system, the current sheet moves from the polar region back and forth across the equator, which warps the current sheet (see in Figure \ref{fig:panels} the second row the second and third panel). 
The oscillating motion is dampened over time, until ultimately after $\sim 60 \ \gls{gravradius} / \gls{spol}$ the magnetosphere settles into a split monopole geometry with a flat current sheet lying in the equatorial plane (see in Figure \ref{fig:panels} in the second row the fourth panel).

\paragraph{Quadrudipole}
The quadrudipole has a dipole moment $\gls{mpolemoment}_{1}$ and quadrupole moment $\gls{mpolemoment}_{2}$ such that $\gls{mpolemoment}_{1} / \gls{mpolemoment}_{2} = (1/30) \gls{gravradius}^{-1}$.
The topology is such that there is one region of one polarity separating two regions of the other polarity.
The addition of the dipole moment breaks the quadrupole symmetry and its symmetric distribution of the regions of ingoing and outgoing flux.
Contrary to the dipole geometry containing poloidal field loops extending over the entire polar angle, for a quadrudipole the largest poloidal field loops extend from the poles to an intermediate polar angle creating two separate poloidal field loop regions.

After poloidal field loops in both regions have fallen into the BH, two conal current sheets form after $\sim 60 \ \gls{gravradius} / \gls{spol}$ (see in Figure \ref{fig:panels} in the third row the second panel and in the right panel the cyan line). 
The current sheets are oriented at an oblique angle with respect to the equatorial plane, therefore in the axisymmetric system they are cone-shaped. 
A region of flux of single polarity that contains one of the poles is enclosed by such a conal current sheet.
Over time, the southern conal current sheet moves towards the southern pole (i.e., the angle at the apex of the southern cone-shaped current sheet decreases) (see in Figure \ref{fig:panels} in the third row the third panel). 
At the same time the northern current sheet moves towards the equator.  
At $\gls{time} \sim 128 \ \gls{gravradius} / \gls{spol}$, the southern conal current sheet narrows in on itself (i.e., the angle at the apex of the southern cone-shaped current sheet becomes zero) and 
vanishes together with the region of flux it enclosed, thereby changing the topology of the magnetic field on the event horizon. 
This leaves the system in a split monopole geometry with a current sheet at the equatorial plane.

\paragraph{Octuquadrupole}
The octuquadrupole with quadrupole moment $\gls{mpolemoment}_{2}$ and octupole moment $\gls{mpolemoment}_{3}$ such that $\gls{mpolemoment}_{2} / \gls{mpolemoment}_{3} = (1/300) \gls{gravradius}^{-1}$ forms three current sheets after $\sim 60 \ \gls{gravradius} / \gls{spol}$ (see in Figure \ref{fig:panels} in the bottom row the first and second panels and in the right panel the black and cyan lines).  
he topology is such that there are two regions of one polarity separating two regions of the other polarity.
The addition of the quadrupole moment breaks the octupole symmetry and its symmetric distribution of regions of ingoing and outgoing flux.

At $\gls{time} \sim 145 \ \gls{gravradius} / \gls{spol}$, the bottom two current sheets merge together causing them to vanish (see in Figure \ref{fig:panels} in the bottom row the third panel and in the right panel the magenta line), making it qualitatively different to the quadrudipole case where one current sheet moved to the south pole. 
As a result, the magnetosphere is left in a split monopole geometry (see in Figure \ref{fig:panels} in the bottom row the fourth panel and in the right panel the red line).

\section{Modelling the magnetospheric dynamics}
The previous qualitative descriptions show that the magnetospheric evolution can be divided into two phases. 
In the first phase, rapid magnetic pressure equilibration drives the plasma on Alfv{\'e}nic timescales towards a pressure balance which is reached for all cases after $\sim 60 \ \gls{gravradius} / \gls{spol}$. 
Herein, poloidal field loops fall into the BH and the initially closed magnetic field geometries open up into (multi-) split monopole magnetospheres, consisting of a purely radial magnetic field with one or more current sheets. 
In the second phase, properly described by our high-resolution simulations, reconnection occurs in the plasmoid-mediated regime in the currents sheets that move (in polar angle) over the horizon. The associated reconnection timescale is much longer than the Alfv{\'e}nic timescale for magnetic pressure equilibration, yet faster than the Ohmic diffusion time scale for small (numerical) resistivity, due to the asymptotic reconnection rate achieved in the plasmoid-mediated regime.
Therefore, we assume that in the second phase pressure equilibrium is continuously maintained such that $| \gls{Bcomp}^{\gls{mksradius}}(\gls{time}) | = \gls{magnflux}_{\textrm{tot}}(\gls{time}) / \gls{area}_{\textrm{tot}} $ remains spatially uniform but decreases over time, which is supported by all our simulations (e.g., see Figure \ref{fig:panels} in the right panels).

We will now develop a quantitative model for the evolution in the second phase.  To do so, we define a flux region to be a region of flux with a single polarity that is either bounded by one current sheet if it contains the pole in an axisymmetric system, or by two current sheets.
For a magnetic field geometry with multipole order $l \in \mathbb{N} $ (e.g., $l = 1$ for the dipole, $l=2$ for the quadrupole, and $l=3$ for the octupole) the total number of flux regions on the event horizon is $\gls{num} = l + 1$.
If $\gls{num}$ is even, the number of flux regions of each polarity are equal and if $\gls{num}$ is odd, one polarity has one more flux region than the other.
In order to obtain pressure equilibrium, the flux $\gls{magnflux}_{n}(\gls{time})$ through flux region $n$ must be directly proportional to the surface area of that region $\gls{area}_{n} (\gls{time})$ on the event horizon. 
In other words, this means that in order to maintain pressure balance the fraction of $\gls{magnflux}_{n}(\gls{time})$ relative to the total event horizon flux $ \gls{magnflux}_{\textrm{tot}} (\gls{time})= \sum_{n=1}^{N}  \gls{magnflux}_{n} (\gls{time})$ must be equal to the surface area fraction $\gls{areafrac}_{n}(\gls{time}) = \gls{area}_{n}(\gls{time}) / \gls{area}_{\textrm{tot}}$  of that region relative to the total event horizon surface area $\gls{area}_{\textrm{tot}} $ giving, 
\begin{eqnarray}
    \gls{areafrac}_{n}(\gls{time})
    =
    \frac{ \gls{magnflux}_{n} (\gls{time}) }{  \gls{magnflux}_{\textrm{tot}} (\gls{time})}   \hspace{0.5cm}, n \in \{1,\dots,N\}
\end{eqnarray}
The current sheets on the flux region boundaries in terms of the normalized cumulative surface coordinate $\gls{cumsurfcoord}$ are then located at,
\begin{eqnarray}
    \gls{cumsurfcoord}_{\textrm{cs},j}(\gls{time})
    =
    \frac{\sum\limits_{n=1}^{j}  \gls{magnflux}_{n} (\gls{time}) }{  \gls{magnflux}_{\textrm{tot}} (\gls{time})}   \hspace{0.5cm}, j \in \{1,\dots,N-1\}
\end{eqnarray}

For a surface lying on the event horizon ($\gls{sphradius} = \gls{ehradius} = 2 \gls{gravradius}$) with a bounding contour on the current sheet the flux decay through reconnection on one side of a current sheet can be determined to be 
$
\partial_{\gls{time}} \gls{magnflux}   
= 
- 2 \pi \gls{lapse}    \sqrt{\gls{spmetricdet}} 
(\gls{3velcomp}^{\widehat{\gls{sphpolar}}} /  \sqrt{\gls{spmetriccomp}_{\gls{sphpolar}\gls{sphpolar}}} ) 
(\gls{magnflux}_{\textrm{tot}}  / \gls{area}_{\textrm{tot}} ) 
= 
- \gls{lapse} \gls{3velcomp}^{\widehat{\gls{sphpolar}}}  \gls{magnflux}_{\textrm{tot}} \ \textrm{sin}(\gls{sphpolar}_{\textrm{cs}}) / ( 4 \gls{gravradius} ) 
$.
Here,  
$\gls{lapse}$ is the lapse function, 
$\gls{sphpolar}_{\textrm{cs}}$ is the polar-angle of the current sheet location, 
$\gls{3velcomp}^{\widehat{   i }} = \sqrt{ \gls{spmetriccomp}_{ii}} \gls{3velcomp}^{i}  $ is the physical 3-velocity, 
$\gls{3velcomp}^{i}$ is the coordinate 3-velocity, 
$\gls{spmetriccomp}_{ii}$ are the components of the spatial part of the metric, and
for a non-spinning BH $\gls{spmetricdet} = \gls{spmetriccomp}_{\gls{mksradius}\gls{mksradius}} \gls{spmetriccomp}_{\gls{sphpolar}\gls{sphpolar}} \gls{spmetriccomp}_{\gls{sphazi}\gls{sphazi}}$.

Here, the factor $\textrm{sin}(\gls{sphpolar}_{\textrm{cs}})$ accounts for differences in the circumference of the reconnection layer on the event horizon for different latitudes.
Magnetic flux decay occurs at the same rate on both sides of a current sheet. 
Therefore, an equal amount of flux of each polarity is reconnected per unit of time such that the total magnetic flux (i.e., not the absolute magnetic flux) through the entire BH event horizon (i.e., a closed surface) remains zero. 

For any axisymmetric magnetic field geometry only the two flux regions containing the poles are bounded by a single current sheet.
The other flux regions that lie in between are bounded by two current sheets.
Hence the flux evolution can be described by the set of coupled ordinary differential equations (ODEs),
\begin{eqnarray}
    \frac{\partial \gls{magnflux}_{1}}{\partial \gls{time}}
    &&=
    - \frac{\gls{lapse} \gls{3velcomp}^{\widehat{\gls{sphpolar}}}  }{4 \gls{gravradius}}
    \gls{magnflux}_{\textrm{tot}} 
    \ \textrm{sin}(\gls{sphpolar}_{\textrm{cs},1}) 
    \nonumber \\
    \frac{\partial \gls{magnflux}_{2}}{\partial \gls{time}}
    &&=
    - \frac{\gls{lapse} \gls{3velcomp}^{\widehat{\gls{sphpolar}}}  }{4 \gls{gravradius}}
    \gls{magnflux}_{\textrm{tot}} 
    \ (\textrm{sin}(\gls{sphpolar}_{\textrm{cs},1}) + \textrm{sin}(\gls{sphpolar}_{\textrm{cs},2}) )
    \nonumber \\
    &&\vdots
    \nonumber \\
    \frac{\partial \gls{magnflux}_{N-1}}{\partial \gls{time}}
    &&=
    - \frac{\gls{lapse} \gls{3velcomp}^{\widehat{\gls{sphpolar}}}  }{4 \gls{gravradius}}
    \gls{magnflux}_{\textrm{tot}} 
    \ (\textrm{sin}(\gls{sphpolar}_{\textrm{cs},N-2}) + \textrm{sin}(\gls{sphpolar}_{\textrm{cs},N-1}) )  
    \nonumber \\
    \frac{\partial \gls{magnflux}_{N}}{\partial \gls{time}}
    &&=
    - \frac{\gls{lapse} \gls{3velcomp}^{\widehat{\gls{sphpolar}}}  }{4 \gls{gravradius}}
    \gls{magnflux}_{\textrm{tot}} 
    \ \textrm{sin}(\gls{sphpolar}_{\textrm{cs},N-1}) 
\label{eq:odes}
\end{eqnarray} 
with 
$\textrm{sin}(\gls{sphpolar}) = \sqrt{1- (1-2 \gls{cumsurfcoord})^{2}}$.
By solving this set of ODEs (\ref{eq:odes}) for initial conditions taken in a pressure equilibrium state, we can predict the evolution of the individual fluxes $\gls{magnflux}_{n} (\gls{time})$, the total flux $ \gls{magnflux}_{\textrm{tot}} (\gls{time})$, the surface area fractions $\gls{areafrac}_{n}(\gls{time})$, and the evolution of current sheet locations $\gls{cumsurfcoord}_{\textrm{cs},j}(\gls{time})$ (or equivalently $\gls{sphpolar}_{\textrm{cs},j}(\gls{time})(\gls{cumsurfcoord}_{\textrm{cs},j})$ with $\gls{sphpolar}(\gls{cumsurfcoord}) = \textrm{arccos}(1-2 \gls{cumsurfcoord})$).

\paragraph{Predicting the magnetospheric evolution}
We now use this analytic model to explain the magnetospheric evolution of our simulated cases. 

\begin{figure*}[ht!]
\centering
\resizebox{\textwidth}{!}{ 
\plotone{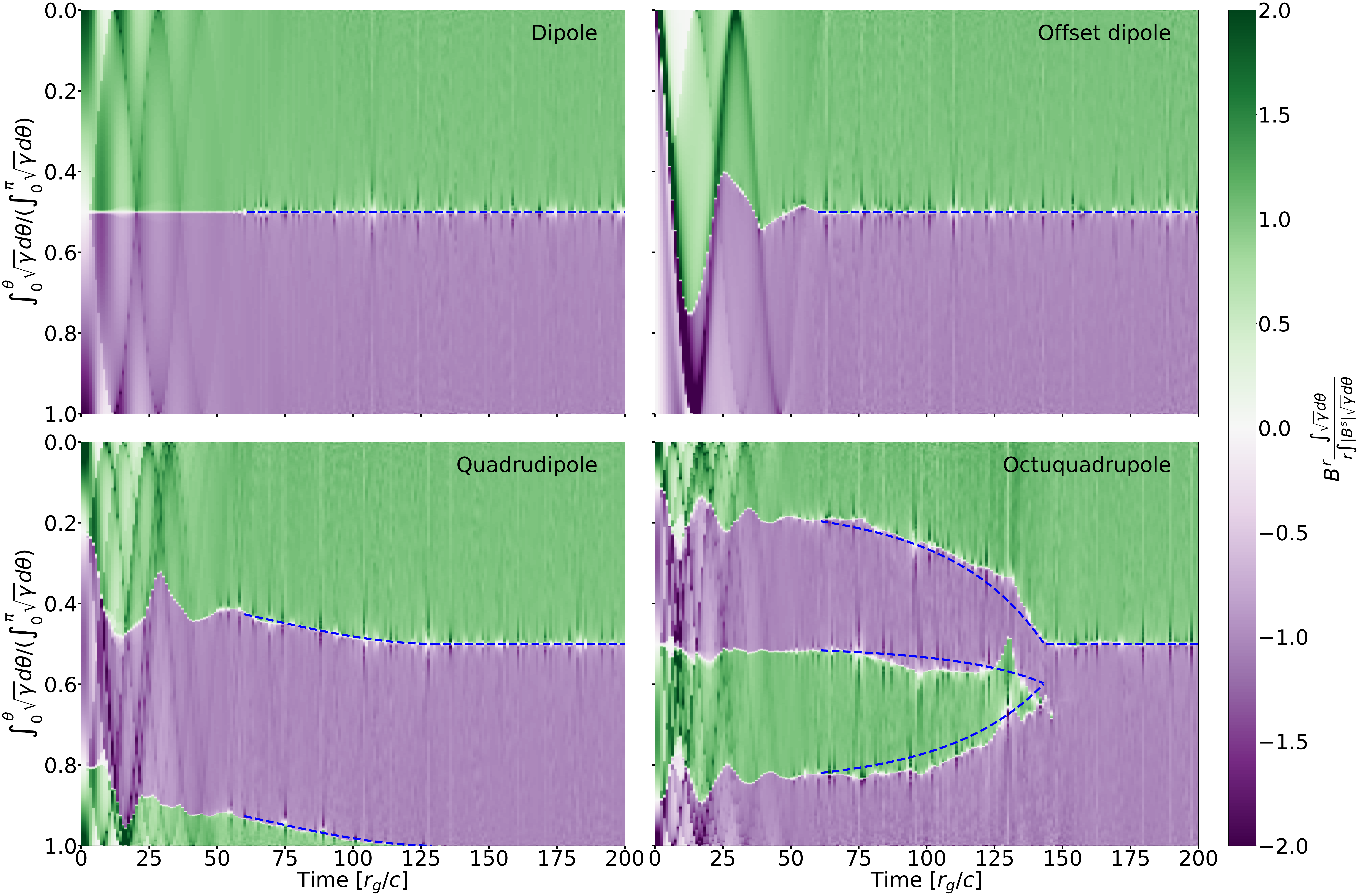}
}
\caption{   
Spacetime plots for the 4 initial magnetic field geometries of the normalized radial magnetic field over the event horizon as a function of cumulative surface coordinate $\gls{cumsurfcoord}$ along the vertical axis and as a function of time along the horizontal axis. The two colors indicate the surface area of flux regions of different polarity and the current sheet is located at the white interface where $\gls{Bcomp}^{\gls{sphradius}}$ changes sign. The blue dashed lines are the current sheet locations predicted from solving the set of ODEs (\ref{eq:odes}) using the initial conditions for each simulation at $\gls{time} = 60 \ \gls{gravradius} / \gls{spol}$ when pressure equilibrium is established.
}
\label{fig:spacetimeplots}
\end{figure*}

Figure \ref{fig:spacetimeplots} displays for the four initial magnetic field geometries a spacetime plot of the normalized radial magnetic field (equivalent to the quantity on the vertical axis in the right column of Figure \ref{fig:panels}) at the event horizon as a function of $\gls{cumsurfcoord}$ along the vertical axis and as a function of time along the horizontal axis, as measured in our simulations.
The initial phase of pressure equilibration is accompanied by waves propagating through the system showing up in the spacetime plots as sinusoidal patterns in the radial magnetic field strength. 
This Figure \ref{fig:spacetimeplots} demonstrates that for all cases pressure equilibrium is established around $\gls{time} \sim 60 \ \gls{gravradius} / \gls{spol}$ as the sinusoidal patterns vanish and the radial magnetic field strength becomes nearly uniform. 

For each case, the state of the system at $\gls{time} = 60  \ \gls{gravradius} / \gls{spol}$ serves as the initial condition for the set of ODEs (\ref{eq:odes}). 
Furthermore, from our dipole simulation we measure at the event horizon $\gls{sphradius} = 2 \gls{gravradius}$ a polar velocity of inflow into the current sheet of $\gls{3velcomp}^{\widehat{\gls{sphpolar}}}=0.028 \gls{spol}$, which is roughly consistent with the expected $\sim 0.01 \gls{spol}$ for plasmoid-mediated reconnection \citep{Loureiro2007} as well as with resistive GRMHD simulations of BH balding, indeed showing a slightly larger inflow velocity $\sim 0.03 \gls{spol}$ \citep{Bransgrove2021}. 
We numerically solve the set of ODEs for each case, allowing us to predict the evolution of the current sheet locations to good agreement with our simulation results (in Figure \ref{fig:spacetimeplots} compare the blue dashed lines of our model to the white region of the colormap indicating the current sheet position).
If during the magnetospheric evolution one of the flux regions disappears the magnetic field topology on the event horizon changes and the validity of the predicted evolution of the solved set of ODEs stops. 
The state of the system at that time then serves as the initial condition for a new set of ODEs that is applicable to the newly formed topology (having one fewer flux region). 
This new set of ODEs can be solved to predict the magnetospheric evolution from then on.

We can now describe each simulated case in more detail. 
For the initial dipole (see in Figure \ref{fig:spacetimeplots} the top left panel), its initial geometry has a pressure imbalance along the event horizon (i.e., a $\gls{sphpolar}$-dependence of the magnetic pressure). 
Both infalling flux and magnetic pressure imbalance drive magnetic pressure equilibration to move flux from the polar regions to the equatorial region.
The initial dipole has two flux regions, which maintain their initially equal surface area and the current sheet position remains at the equator once the split monopole has formed.

For the initial offset dipole, the magnetic pressure in the smaller flux region (i.e. the region with smaller surface area, not with less flux) is higher than in the larger flux region (see in Figure \ref{fig:panels} in the first row the first panel and in the right panel the black line and in Figure \ref{fig:spacetimeplots} the top right panel).
Pressure equilibration forces a surface area increase of the smaller flux region and an inevitable surface area decrease of the larger flux region. Ultimately, in pressure balance both regions have equal surface areas forming a split monopole geometry.

The topology of the quadrudipole consists of two flux regions with the same polarity each containing one of the poles separated by one flux region of the opposite polarity in between (see in Figure \ref{fig:panels} the third row the first panel and in Figure \ref{fig:spacetimeplots} the bottom left panel).
The northern flux region contains more flux than the southern one with the same polarity.
This asymmetry is amplified during the pressure equilibration phase (see in Figure \ref{fig:panels} in the third row the first panel and in the right panel the black line). 
In the second phase of pressure balance, the fraction of flux of the southern flux region relative to the total flux decreases, which leads to a decrease of its surface area and moves the bounding southern current sheet toward the south pole.
When the southern flux region runs out of flux at $\gls{time} =  128 \ \gls{gravradius} / \gls{spol}$, the conal current sheet narrows in on itself and vanishes, leaving a split monopole geometry.

The octuquadrupole has four flux regions which after pressure equilibration are bounded by three current sheets  (see in Figure \ref{fig:panels} in the bottom row the first and second and in the right panel the black and cyan line and in Figure \ref{fig:spacetimeplots} the bottom right panel).
The region bounded by the bottom two current sheets runs out of flux first, thereby merging two current sheets and two flux regions of the same polarity, in agreement with the analytical predictions of our model.
This leaves the magnetosphere in a split monopole geometry (see in Figure \ref{fig:panels} in the bottom row the third and fourth panel and in the right panel the magenta and red line).

\paragraph{Quadrupole}
In addition to the four initial magnetic field geometries considered until now, we performed a simulation with an initial quadrupole geometry. 
The initial topology on the event horizon of the quadrupole is equal to that of the quadrudipole. 
However, contrary to the quadrudipole, the quadrupole is symmetric as both the northern and southern flux region with the same polarity have an equal amount of flux. 
This symmetry is largely conserved during the pressure equilibration phase lasting, similarly to all other simulated cases, for $\sim 60 \ \gls{gravradius} / \gls{spol}$. 
During pressure equilibration two conal current sheets form that have a comparable angle with respect to the equatorial plane.
In the phase of pressure balance, initially the flux decay in both current sheets is equal such that the flux in the two regions of the same polarity remains equal, thereby maintaining the symmetry of the system. 
However, stochastic plasmoid formation in both current sheets over time leads to a symmetry breaking in the flux decay rate between the two current sheets. 
As a result, the symmetry between the flux regions of the pure quadrupole is broken and the system starts to evolve qualitatively similar to the previously described quadrudipole. 
Ultimately, one of the current sheets narrows in on itself and vanishes leaving the system in a split monopole geometry.

While our analytic model does not account for stochastic plasmoid formation and thus preserves the symmetry of the flux regions indefinitely, this is not borne out in the simulations. 
Our numerical experiments show that the time it takes for the symmetry to break, which leads to a split monopole geometry is $\sim 300 \ \gls{gravradius} / \gls{spol}$. 
Hence we conclude that $\sim 300 \ \gls{gravradius} / \gls{spol}$ is an upper limit to the formation of a split monopole for the initial quadrupolar field. 
The symmetry breaking in the simulations is seeded by numerical noise which amplifies in the chaotic evolution of the plasmoid chain.

\section{Magnetic flux decay}
\label{sec:magnfluxdecay}
In this section we discuss in more detail the evolution of the magnetic flux. 
Inside a current sheet magnetic reconnection dissipates magnetic field energy that is transferred to the thermal and kinetic energy of the plasma. 
Dissipation causes a decrease of the radial magnetic field strength leading to event horizon magnetic flux decay.  

\begin{figure}[ht!]
\centering
\resizebox{\columnwidth}{!}{ 
\plotone{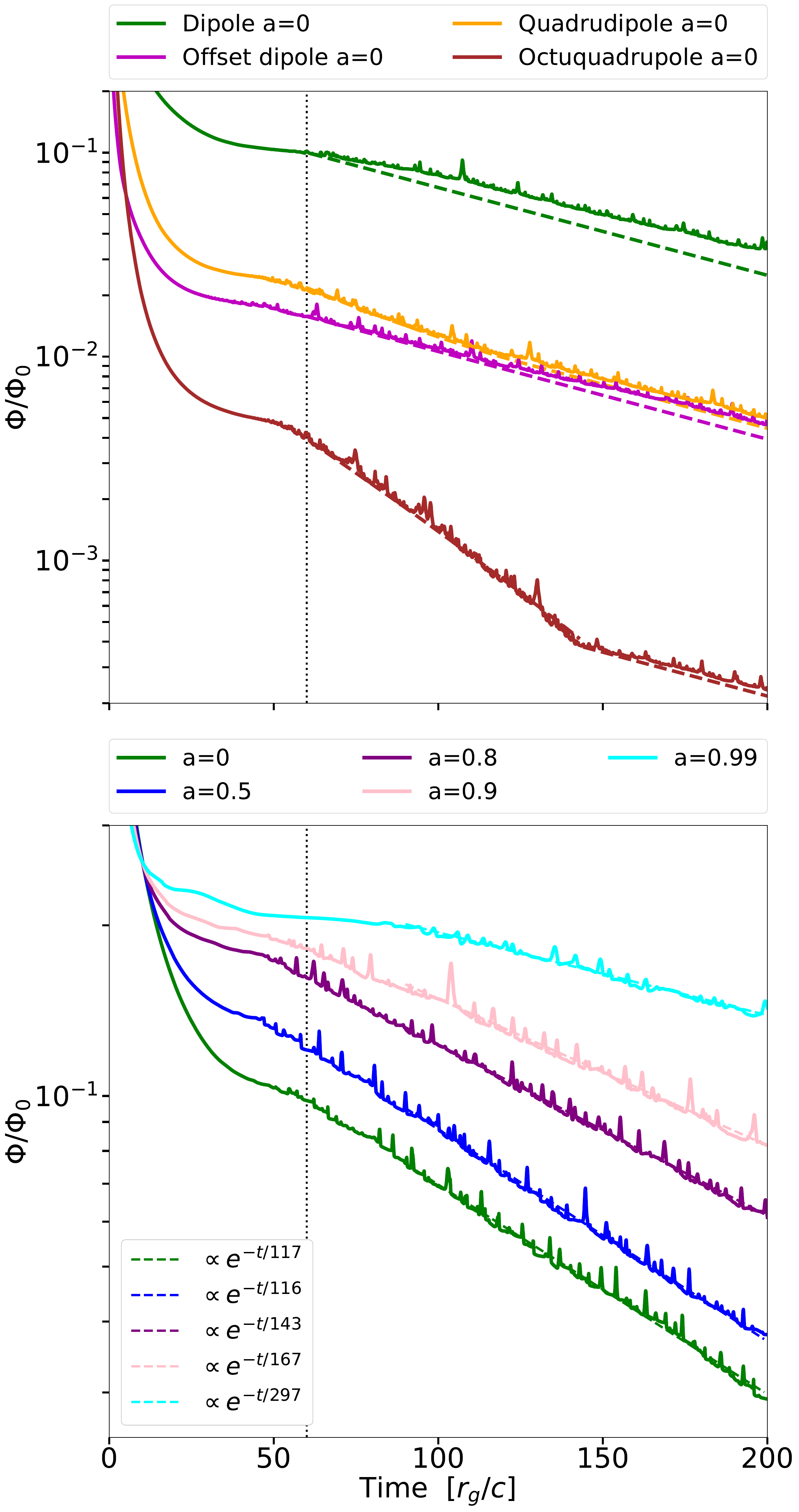}
}
\caption{   
The total absolute flux $\gls{magnflux}_{\textrm{tot}} (\gls{time})$ at event horizon normalized by the initial total event horizon absolute flux $\gls{magnflux}_{0}= \gls{magnflux} (\gls{time}=0)$ as a function of time. 
The top panel shows four simulated initial magnetic field geometries and their predicted evolution by the analytic model, shown by respectively the colored continuous lines and the corresponding colored dashed lines. 
The lower panel shows five simulations with an initial dipole for five different BH spins and their fits with an exponential function in the second pressure equilibrium evolutionary phase, shown by respectively the colored continuous lines and the corresponding colored dashed lines. 
The vertical dotted line in both panels shows the time $60 \ \gls{gravradius} / \gls{spol}$ at which pressure equilibrium is established for all simulations, thereby separating the first pressure equilibration phase from the second pressure balance phase.
}
\label{fig:fluxdecay}
\end{figure}

\paragraph{Magnetic flux decay during magnetospheric evolution}
In Figure \ref{fig:fluxdecay}, the top panel displays for the four simulated initial magnetic field geometries of section \ref{sec:magnetopshericevolution} the time evolution of $\gls{magnflux}_{\textrm{tot}} (\gls{time})$ normalized by the initial total event horizon absolute flux $\gls{magnflux}_{0}= \gls{magnflux}_{\textrm{tot}} (\gls{time}=0)$.

During the pressure equilibration phase $\gls{time} \lesssim 60 \ \gls{gravradius} / \gls{spol}$, the poloidal field loops falling into the BH lead to a large drop in $\gls{magnflux}_{\textrm{tot}} (\gls{time})$. 
The magnitude of the flux drop depends on the initial magnetic field geometry
as flux moves over the BH event horizon (see in Figure \ref{fig:panels} the leftmost three columns).
The smallest simulated flux drop to $\sim 0.1\gls{magnflux}_{0}$ is for the initial dipole and the largest to $\sim 0.005\gls{magnflux}_{0}$ is for the octuquadrupole.
During this first phase the time-varying magnetic field on the event horizon is accompanied by an azimuthal electric field $\gls{cEcomp}_{\gls{sphazi}} =  \sqrt{\gls{spmetricdet}} ( \gls{Bcomp}^{\gls{mksradius}}  \gls{trvelcomp}^{\gls{sphpolar}}  -  \gls{Bcomp}^{\gls{sphpolar}}  \gls{trvelcomp}^{\gls{mksradius}} )  / \gls{spol} $  with $\gls{trvelcomp}^{i}= \gls{lapse} \gls{3velcomp}^{i} - \gls{shiftv}^{i}$ the transport velocity and $\gls{shiftv}^{i}$ the shift vector.
Here, the flux drop is dominated by the second term in the electric field, which initially is much larger than the first term, as infalling poloidal field loops make $\gls{Bcomp}^{\gls{sphpolar}}$ cross the event horizon with $ \gls{trvelcomp}^{\gls{mksradius}} \gtrsim 0.1 \gls{spol}$. 

In the second phase of pressure balance, in addition to the current sheet positions, our analytic model (\ref{eq:odes}) allows us to predict magnetic flux decay on the event horizon. 
Once any initial geometry has transitioned to the split monopole geometry, the set of ODEs (\ref{eq:odes}) can even be solved analytically for the evolution of the total event horizon flux. 
In the split monopole phase, the flux is predicted to decay exponentially with exponential decay timescale $\gls{tscale} = \gls{sphradius}_{\textrm{H}}  / ( \gls{lapse} \gls{3velcomp}^{\widehat{\gls{sphpolar}}} ) \approx 101 \ \gls{gravradius} / \gls{spol}$ for a non-spinning BH. This is in good agreement with the decay measured in all simulations once the system enters the split monopole phase (see in Figure \ref{fig:fluxdecay} in the top panel).

The offset dipole has an initial flux drop down to $\sim 0.02\gls{magnflux}_{0}$ after which in the second phase it decays with an equivalent timescale as the initial dipole case (see in Figure \ref{fig:fluxdecay} in the top panel the magenta continuous line and the magenta dashed line).
For the quadrudipole, after the initial flux drop down to $\sim 0.03\gls{magnflux}_{0}$, two current sheets contribute to the decay rate resulting in a faster decay compared to cases with a single current sheet (see in Figure \ref{fig:fluxdecay} in the top panel the orange continuous line and the orange dashed line). 
As one of the current sheets moves to the coordinate pole, its decreasing circumference around the event horizon decreases its contribution to the total flux decay rate. 
This continues until the conical current sheet narrows in on itself and vanishes at the pole at $\gls{time} \sim 128  \ \gls{gravradius} / \gls{spol}$ and the flux decay seamlessly goes into the exponential decay of the remaining split monopole, with a decay rate similar to both dipole cases.
For the octuquadrupole the three current sheets cause the fastest flux decay rate of all simulated cases (see in Figure \ref{fig:fluxdecay} in the top panel the brown continuous line and the brown dashed line). 
The merging of the two current sheets at $\gls{time} \sim 145 \ \gls{gravradius} / \gls{spol}$ occurs not near the coordinate poles but in the equatorial region. Therefore, before merger, both current sheets had a large contribution to the total flux decay. 
As a result of the two current sheets merging and vanishing, the decay rate after that time drops significantly (see in Figure \ref{fig:fluxdecay} the inflection point in the brown lines) to the split monopolar exponential decay rate.

In addition to magnetic flux decay through the decrease of magnetic field strength, magnetic flux can also move across the event horizon. 
This can occur through plasmoids falling into the BH which will first increase the event horizon flux when the plasmoid starts crossing the event horizon and will decrease once the null point of the plasmoid center crosses the event horizon. 
This effect shows up in the magnetic flux decay as small spikes on top of the main magnetic flux decay.

\paragraph{Spin dependence of magnetic flux decay} 
We now consider the spin dependence of the magnetic flux decay on the event horizon. 
For this we use an initial dipole geometry for a range of dimensionless BH spins $\gls{spinpar} \in \{0,0.5,0.8,0.9,0.99\}$, which all evolve to a split monopole geometry.
In Figure \ref{fig:fluxdecay} the bottom panel displays the time evolution of the normalized event horizon flux $\gls{magnflux}_{\textrm{tot}} (\gls{time}) / \gls{magnflux}_{0}$ for the five simulated spins.  
Similar to the non-spinning cases, for all spins there is an initial pressure equilibration phase $\gls{time} \lesssim 60 \ \gls{gravradius} / \gls{spol}$ accompanied by a flux drop dominated by poloidal field falling into the BH.
The flux drops for the lowest spin $\gls{spinpar}=0$ to $\sim 0.1\gls{magnflux}_{0}$ and for the highest simulated spin $\gls{spinpar}=0.99$ to $\sim 0.2 \gls{magnflux}_{0}$.

Due to spacetime rotation, the split monopole field redistributes into a non-uniform polar-angle profile over the event horizon. 
More specifically, the magnitude of the radial magnetic field $|\gls{Bcomp}^{\gls{mksradius}}|$ peaks at the rotational poles and is lowest at the rotational equator.
We quantify this non-uniformity similar to \citet{Bransgrove2021} by expressing $\gls{magnflux}_{\textrm{tot}} = k \gls{area}_{\textrm{tot}} \gls{Bcomp}^{\gls{mksradius}}(\gls{sphpolar} = \pi/2) $.
This defines a non-uniformity factor $k$, where $k=1$ represents a uniform distribution and for $k>1$ more flux is concentrated at the poles than at the equator. 
For the simulated spins we measure $k \in \{1, 1.04, 1.13, 1.17, 1.56\}$, where a higher spin corresponds to a higher degree of non-uniformity.

For all spins the flux decays exponentially in the split monopolar phase, where higher spin results in slower decay.
Fits with an exponential function $\propto e^{-\gls{time} / \tau}$ (see in Figure \ref{fig:fluxdecay} in the bottom panel the dashed lines with colors corresponding to the colored lines of the simulations) show that the decay timescale $\tau$ ranges from $\sim 117 \ \gls{gravradius} / \gls{spol}$ for $\gls{spinpar}=0$ to $\sim 297 \ \gls{gravradius} / \gls{spol}$ for $\gls{spinpar} = 0.99$.
Using the expression for $\gls{magnflux}_{\textrm{tot}}$ in terms of $k$ the exponential decay timescale is $
\tau 
=
k \gls{area}_{\textrm{tot}}  \sqrt{\gls{spmetriccomp}_{\gls{sphpolar}\gls{sphpolar}}}   
/ 
(4 \pi \gls{lapse}  \sqrt{\gls{spmetricdet}} \gls{3velcomp}^{\widehat{\gls{sphpolar}}}   )
$. 
Evaluating the analytic quantities $\gls{area}_{\textrm{tot}}$, $\gls{spmetriccomp}_{\gls{sphpolar}\gls{sphpolar}}$, $\gls{lapse}$, and $\gls{spmetricdet}$ at the BH event horizon at the equatorial plane (i.e., $(\gls{sphradius},\gls{sphpolar}) = (\gls{ehradius},\pi/2)$) and using the measured $k$ and $\gls{tscale}$ for all spins sets the inflow velocities $\gls{3velcomp}^{\widehat{\gls{sphpolar}}}(\gls{sphradius}=\gls{ehradius},\gls{sphpolar} = \pi/2) \in \{0.024, 0.025, 0.020, 0.017, 0.016\}$ into the equatorial current sheet. 
This suggests that the inflow velocity depends on the BH spin, where higher spin results in a lower inflow velocity. 
The measured trend is similar to the one analytically determined by \citet{Asenjo2017}, however that model is only valid far away from the event horizon and for low spin.

\section{Discussion}

\paragraph{Energetics}
We provide here some basic energetic estimates. 
Assume that the magnetic flux $\gls{magnflux}_{0}$ equal to that through the surface of a star with a polar magnetic field strength $\gls{Bcomp}_{\textrm{p}}$ and radius $\gls{radius}_{*}$ is transferred to a BH with mass $\gls{mass}_{\textrm{BH}}$ as a dipolar magnetic field. 
The magnetic field evolves from a dipole to a split monopole during which the flux through event horizon drops to $\gls{factor}\gls{magnflux}_{0}$. 
Here, the factor $\gls{factor}$ depends on the considered scenario as shown in our simulations (see Figure \ref{fig:fluxdecay}).

The initial offset dipole (in Figure \ref{fig:fluxdecay} in the top panel the magenta line) may serve as a simplified model of a merger remnant of a BH with a (neutron/white dwarf) star (i.e., BH-(N/WD)S) at the time the star has crossed over the event horizon. 
This case leads to a flux drop with $\gls{factor} \sim 0.02$ for a non-spinning BH regardless of the BH mass. 
\citet{Kim2024} showed that for a $1.8 \gls{solmass}$ NS merging with a $8 \gls{solmass}$ BH with spin $\gls{spinpar}=0.3$ the split monopole field inherits $\sim 1 \%$ of the magnetic flux from the companion NS, which is similar to our results for a non-spinning BH. 

The centered dipole (in Figure \ref{fig:fluxdecay} in the bottom panel) may serve as a simplified model for a NS-NS merger remnant or a collapsed NS leading to $\gls{factor} \sim 0.1-0.2$ depending on the BH spin. 
In this case the BH mass is, in principle, bounded on the lower end by a gravitationally collapsing maximum-mass non-rotating NS \citep{Rezzolla2018} and on the higher end by the (unlikely) merger of two supra-massive NSs \citep{Falcke2014} that are each maximally spinning, such that the range of masses we consider is $2.0 \lesssim \gls{mass}   / \gls{mass}_{\odot} \lesssim 4.8$.

In the split monopole, magnetic energy dissipates through magnetic reconnection in the equatorial current sheet and is transferred to the plasma (i.e., heating/acceleration of plasma particles). 
The flux decay on the event horizon is $\gls{magnflux}(\gls{time})= \gls{factor} \gls{magnflux}_{0} e^{-\gls{time}/\gls{tscale}} $ with $\gls{tscale} $ the decay timescale. 
The (upstream) split monopolar radial magnetic field strength $\gls{Bcomp}_{\textrm{sm}}$ is proportional to the inverse square of the radial distance and in the case of a non-spinning BH is given at the event horizon by expression (\ref{eq:bmono}).
The dissipated power $\gls{power}_{\textrm{B}}$ in the non-spinning case of the total magnetic field \footnote{We use the dissipated power in the entire magnetic field from the event horizon to infinity. For $\gls{spinpar}=0$, however, 90\% of the total magnetic field energy is contained within $\gls{sphradius} < 20 \gls{gravradius}$.} is given by expression (\ref{eq:powerdiss}).
The dissipation power yields an upper limit for the available energy for emission through magnetic reconnection.
For spinning BHs, the extracted power $\gls{power}_{\textrm{BZ}}$ from the BH spin via the Blandford-Znajek (BZ) process \citep{Blandford1977,Tchekhovskoy2010a} is given by expression (\ref{eq:powerbz}) with $\kappa = 1/(6 \pi) $ for a monopolar field.
\begin{widetext}
\begin{eqnarray}
    \gls{Bcomp}_{\textrm{sm}} (\gls{time})
    =
    1.1 \cdot 10^{9}
    \left( \frac{\gls{factor}}{0.02} \right) 
    \left( \frac{\gls{Bcomp}_{\textrm{p}} \gls{radius}_{*}^2}{10^{12} \ \textrm{Gauss} \  \textrm{km}^{2}}\right) 
    \left( \frac{\gls{mass}_{\textrm{BH}}}{\gls{solmass}}\right)^{-2}
    e^{-\frac{  \gls{time}}{ \gls{tscale}}} 
    \ \textrm{Gauss}
    \label{eq:bmono}
\end{eqnarray}
\begin{eqnarray}
    \gls{power}_{\textrm{B}} (\gls{time})
    =
    6.8 \cdot 10^{37}
    \left( \frac{\gls{factor}}{0.02} \right)^{2} 
     \left( \frac{\gls{Bcomp}_{\textrm{p}} \gls{radius}_{*}^2}{10^{12} \ \textrm{Gauss} \  \textrm{km}^{2}}\right)^{2}
     \left( \frac{\gls{spol} \gls{tscale} }{100 \gls{gravradius}} \right)^{-1}
     \left( \frac{\gls{mass}_{\textrm{BH}}}{\gls{solmass}}\right)^{-2}
    e^{-\frac{  2 \gls{time}}{ \gls{tscale}}}
    \ \textrm{erg} \ \textrm{s}^{-1} 
\label{eq:powerdiss}
\end{eqnarray}
\begin{eqnarray}
    \gls{power}_{\textrm{BZ}} (\gls{time})
    =
    2.9 \cdot 10^{40}
    \left( \frac{\kappa}{1 / (6 \pi )} \right)
    \left( \frac{\gls{factor}}{0.02} \right)^{2}
    \left( \frac{\gls{Bcomp}_{\textrm{p}} \gls{radius}_{*}^2}{10^{12} \ \textrm{Gauss} \  \textrm{km}^{2}}\right)^{2}
    \frac{\gls{spinpar}^2}{(1+\sqrt{1-\gls{spinpar}^2})^2}
    \left( \frac{\gls{mass}_{\textrm{BH}}}{\gls{solmass}}\right)^{-2}
    e^{-\frac{2t}{\gls{tscale}}}
    \  \textrm{erg} \ \textrm{s}^{-1} 
\label{eq:powerbz}
\end{eqnarray}
\end{widetext}

Consider a non-spinning BH in the BH-(N/WD)S merger remnant scenario with a flux drop $\gls{factor} = 0.02$ and $\gls{tscale} = 100 \ \gls{gravradius} / \gls{spol}$ acquiring the magnetic flux of a star having $\gls{Bcomp}_{\textrm{p}} \gls{radius}_{*}^{2} = 10^{12} \ \textrm{Gauss} \ \textrm{km}^{2}$ such as a magnetized NS, magnetized WD star, or a main sequence star (e.g., the Sun).
For example, a $10 \gls{solmass}$ stellar mass BH has
$
\gls{Bcomp}_{\textrm{sm}} (\gls{time}) 
\sim 
1.1 \cdot 10^{7} 
e^{-\gls{spol} \gls{time}/100 \gls{gravradius}}  
\ 
\textrm{Gauss}
$ and 
$\gls{power}_{\textrm{B}} (\gls{time}) \sim 6.8 \cdot 10^{35}  e^{-\gls{spol} \gls{time}/50 \gls{gravradius}}  \ \textrm{erg} \  \textrm{s}^{-1}$ 
exponentially decaying on timescales of respectively $\sim 5 \ \textrm{ms}$ and $ \sim 2.5 \ \textrm{ms}$ with a total energy release \footnote{This is the total energy released from $\gls{time} = 0 $ to $\gls{time} \to \infty$.} 
of $1.7 \cdot 10^{33} \ \textrm{erg}$. 
On the other hand, a $10^{6} \gls{solmass}$ supermassive BH has $
\gls{Bcomp}_{\textrm{sm}} (\gls{time}) 
\sim 
1.1 \cdot 10^{-3} 
e^{-\gls{spol} \gls{time}/100 \gls{gravradius}}  
\ 
\textrm{Gauss}
$ and 
$\gls{power}_{\textrm{B}} (\gls{time}) \sim 6.8 \cdot 10^{25}  e^{-\gls{spol}\gls{time}/50\gls{gravradius}}  \ \textrm{erg} \  \textrm{s}^{-1}$ 
exponentially decaying on timescale of respectively $\sim 8 \ \textrm{min}$ and $ \sim 4 \ \textrm{min}$ with a total energy release of $1.7 \cdot 10^{28} \ \textrm{erg}$.

\paragraph{Cooling and emission}
We estimate the cooling regime of reconnection-accelerated particles by comparing the accelerating Lorentz force to the radiation back reaction force and assuming synchrotron cooling dominance. 
Assuming that the BH spin is not too small (i.e., $\gls{spinpar}$ not too close to 0) this results in 
$
(\gls{Bcomp}_{\textrm{sm}}  /\textrm{Gauss})^{3/2}
\lesseqgtr 
2 \gls{mplcty}  / (\gls{mass}_{\textrm{BH}} / \gls{solmass} )
$ 
with $\gls{Bcomp}_{\textrm{sm}}$ the upstream split monopolar magnetic field strength in units of Gauss and $\gls{mplcty}$ the pair multiplicity (i.e., the pair density in excess of Goldreich-Julian). 
In this trichotomy expression, if the left hand side is smaller ($<$) than the right hand side the system is the weak cooling regime, whereas for ($>$) it is in the strong cooling regime, and for ($\sim $) in the intermediate regime.

For the BH-(N/WD)S merger remnant scenario, although $\gls{mplcty}$ is largely unknown, this inequality shows that for a stellar mass BH, reconnection is likely starting in the strong cooling regime, while for a supermassive BH this is less certain to even unlikely. 
As the magnetic field weakens over time the strong cooling regime will go over into the intermediate and ultimately into the weak cooling regime.
The time it takes for this transition to occur is very weakly dependent (i.e., a logarithmic dependence) on all involved quantities other than the decay timescale $\gls{tscale}$. Therefore, the time to transition is approximately a few $\gls{tscale}$.
For a scenario including a stellar mass BH all the magnetic energy dissipated in the current sheet can be expected to immediately be emitted by particles in the fast cooling regime as high-energy (gamma-ray) emission.

For emission to be observable it has to be able to escape the magnetosphere, which may be inhibited by pair production, absorption by merger debris, or surrounding material. 
For stellar mass BHs the fast magnetic flux decay rate would result in fast gamma-ray transients, while for supermassive BHs a transient gamma-ray signal could be observed on timescales similar to minute-scale gamma-ray flares of blazars.

\paragraph{Caveats}
We now comment on the limitations of our work.
In our axisymmetric system we have not taken into account any three-dimensional effects regarding the magnetic field geometry (e.g., \citealt{Selvi2024}) or magnetic reconnection \citep{Zhang2021,Zhang2023,Chernoglazov2023}.
Although realistic BH magnetospheric plasma might be extremely magnetized, we argue that our simulations carried out in the relativistic regime ($\gls{plmagn} \gtrsim 10$) are close to force-free, so that our results are applicable to even more magnetized configurations.
Although our ideal MHD simulations rely on numerical resistivity, we showed that the flux decay timescale has converged (see Appendix \ref{app:numconv}), as reconnection occurs in the plasmoid-mediated regime where the reconnection rate becomes asymptotically independent of resolution, indicating that magnetospheric evolution is not governed by grid-scale dissipation.
Furthermore, the magnetospheric plasma is likely collisionless implying a reconnection rate that is faster by an order of magnitude compared to the collisional reconnection rate observed in MHD simulations \citep{Birn2001,Uzdensky2010,Comisso2016,Cassak2017,Bransgrove2021,Selvi2023,Moran2025}. 
This difference would result in a decrease of the timescale for the magnetic flux decay on the event horizon. 
Another limitation is that we have used a fixed Kerr spacetime. This might not be a serious limitation for the case of a large (e.g., supermassive) BH that acquires the magnetic flux of a much less massive star. However, for nearly equal mass ratio cases, the spacetime evolution will have to be taken into account \citep{Most2019}.

\paragraph{Conclusions}
In this work, we have shown that the absolute magnetic flux through the event horizon of a split monopole magnetosphere decays and depends on spin, with a decay timescale ranging from $\gls{tscale} \sim 117 \ \gls{gravradius} / \gls{spol}$ for $\gls{spinpar}=0$ to $\gls{tscale} \sim 297 \ \gls{gravradius} / \gls{spol}$ for $\gls{spinpar}=0.99$ where higher spin results in slower decay. 
Furthermore, we have shown that for a non-spinning BH the magnetospheric evolution consists of two phases. In the first phase, the magnetosphere evolves toward magnetic pressure equilibrium on a timescale of $\sim 60 \ \gls{gravradius} / \gls{spol}$ for all simulated initial magnetic field geometries. The end state of this first phase is forming a (multi-) split monopole. 
In the second phase, pressure balance is maintained while the magnetic flux decays through magnetic reconnection in current sheets. 
The relative changes of the amount of flux in each flux region result in the (polar angle) movement of current sheet on the event horizon. 
We have presented an analytic model for the second phase describing the evolution of fluxes in each flux region, the total flux, and the movement of current sheets. 
Moreover, we have shown that all simulated initial magnetic field geometries with varying complexity ultimately evolve to a split monopole magnetosphere. 

A BH in the MAD state during the flaring state ($\sim 100 - 200 \ \gls{gravradius} / \gls{spol}$) (i.e, the state in which accretion is momentarily suppressed) \citep{Ripperda2022} may potentially show signs of pressure equilibration, current sheet movement, and current sheet alignment \citep{Selvi2024}.

In a previous work \citep{Selvi2024} we have shown that an initially inclined split monopolar current sheet on a spinning BH over time aligns with the BH equatorial plane. 
Consequently, we argue that for a BH with any spin any acquired magnetic field ultimately evolves to an aligned split monopole magnetosphere.

\section{Acknowledgments}
We would like to thank Yuri Levin, Sasha Philippov and Maxim Lyutikov for insightful discussions and suggestions.
Computational resources and services of this publication were partially provided by the project with No. 2021.001 which is financed by the Dutch Research Council (NWO). 
B.R. is supported by the Natural Sciences \& Engineering Research Council of Canada (NSERC), the Canadian Space Agency (23JWGO2A01), and by a grant from the Simons Foundation (MP-SCMPS-00001470). B.R. acknowledges a guest researcher position at the Flatiron Institute, supported by the Simons Foundation. 
L.S. acknowledges support from DoE Early Career Award DE-SC0023015, NASA ATP 80NSSC24K1238, NASA ATP 80NSSC24K1826, and NSF AST-2307202. 
This work was supported by a grant from the Simons Foundation (MP-SCMPS-00001470) to L.S. and facilitated by Multimessenger Plasma Physics Center (MPPC), grant NSF PHY-2206609 to L.S.
The computational resources and services used in this work were partially provided by facilities supported by the Scientific Computing Core at the Flatiron Institute, a division of the Simons Foundation, and by the VSC (Flemish Supercomputer Center), funded by the Research Foundation Flanders (FWO) and the Flemish Government – department EWI.

\clearpage
\bibliography{references-jabref.bib}{}
\bibliographystyle{aasjournal}

\appendix

\section{Numerical setup initialization and flooring}
\label{app:setup}
In this appendix we provide more details on the initialization and flooring models of the numerical setup.

Initially, the fluid has a vanishing spatial part of the 4-velocity,
$\gls{4velcomp}^i 
= \gls{lfacbulk} (\gls{3velcomp}^{i} - (\gls{shiftv}^{i} / \gls{lapse})) 
= 0$.
 with $\gls{lfacbulk} = (1-\gls{3velcomp}^{2})^{-1/2}$ the Lorentz factor, $\gls{lapse}$ the lapse function, $\gls{3velcomp}^{i}$ the coordinate 3-velocity, and $\gls{shiftv}^{i}$ the shift vector. 
For $\gls{sphradius} \leq 2 \gls{gravradius}$ (i.e., inside a radius of the maximum extent of the ergoregion), the initial plasma magnetization (in cgs units) is set to 
$
\gls{plmagn}(\gls{sphradius} \leq 2 \gls{gravradius}) 
=
d_{0} \gls{sphradius}^{-2} 
=
\gls{bcomp}^{2} /( 4 \pi  \gls{mdens}
\gls{relspecenth})  
$ 
with $d_{0} = 50 \gls{gravradius}^{2}$,
$\gls{relspecenth}
=
( 1 + 
 (\gls{polyidx}/(\gls{polyidx} -1)) 
\gls{dimtemp}  )   \gls{spol}^{2} 
$ the relativistic specific enthalpy, and
$\gls{bcomp}^{i}$ the co-moving magnetic field. 
To avoid that the plasma magnetization becomes so small for large radii that it would no longer represent a force-free magnetosphere we set the initial plasma magnetization for $\gls{sphradius} \geq 2 \gls{gravradius}$ to be uniform and equal to its value at $\gls{sphradius} = 2 \gls{gravradius}$, so
$\gls{plmagn} (\gls{sphradius} \geq 2 \gls{gravradius}) 
%
\sim
12.5$. 
The plasma magnetization $\gls{plmagn} (\gls{sphradius})$ together with the initial magnetic field sets the initial plasma density field $\gls{mdens}(\gls{sphradius},\gls{sphpolar})$,
the initial plasma pressure field $\gls{pres} (\gls{sphradius},\gls{sphpolar}) = \gls{dimtemp} \gls{mdens}(\gls{sphradius},\gls{sphpolar})  \gls{spol}^{2}$, 
and the initial plasma beta (i.e., gas-to-magnetic pressure ratio) 
$\gls{plbeta} (\gls{sphradius})  
=
2 \gls{dimtemp}  \gls{spol}^{2} / (  \gls{plmagn} (\gls{sphradius}) \gls{relspecenth})
$ (e.g., $ \gls{plbeta} (\gls{sphradius} \geq 2 \gls{gravradius}) \sim 6.7 \cdot 10^{-3} $  
).

We employ a floor model for numerical stability in which we enforce a maximum plasma magnetization $\gls{plmagn}_{\textrm{max}}$ with identical radial dependence as the $\gls{plmagn}$ at initialization except with $d_{\textrm{max}} = 100 \gls{gravradius}^{2}$ instead of $d_{0}$ resulting in $\gls{plmagn}_{\textrm{max}} (\gls{sphradius} \geq 2 \gls{gravradius}) = 25$. 
Furthermore, the lack of numerical accuracy in the case of very low density when conserving momentum can lead to unphysical large Lorentz factors. 
In order to diminish the influence of these unphysical cells that can sporadically appear throughout the evolution the Lorentz factor is capped at a maximum of $50$ (see also the discussion how this is handled in modern codes in Section 4.1 of \cite{Porth2019}).

\section{Numerical convergence}
\label{app:numconv}
\begin{figure}[ht!]
\centering
\resizebox{\columnwidth}{!}{ 
\plotone{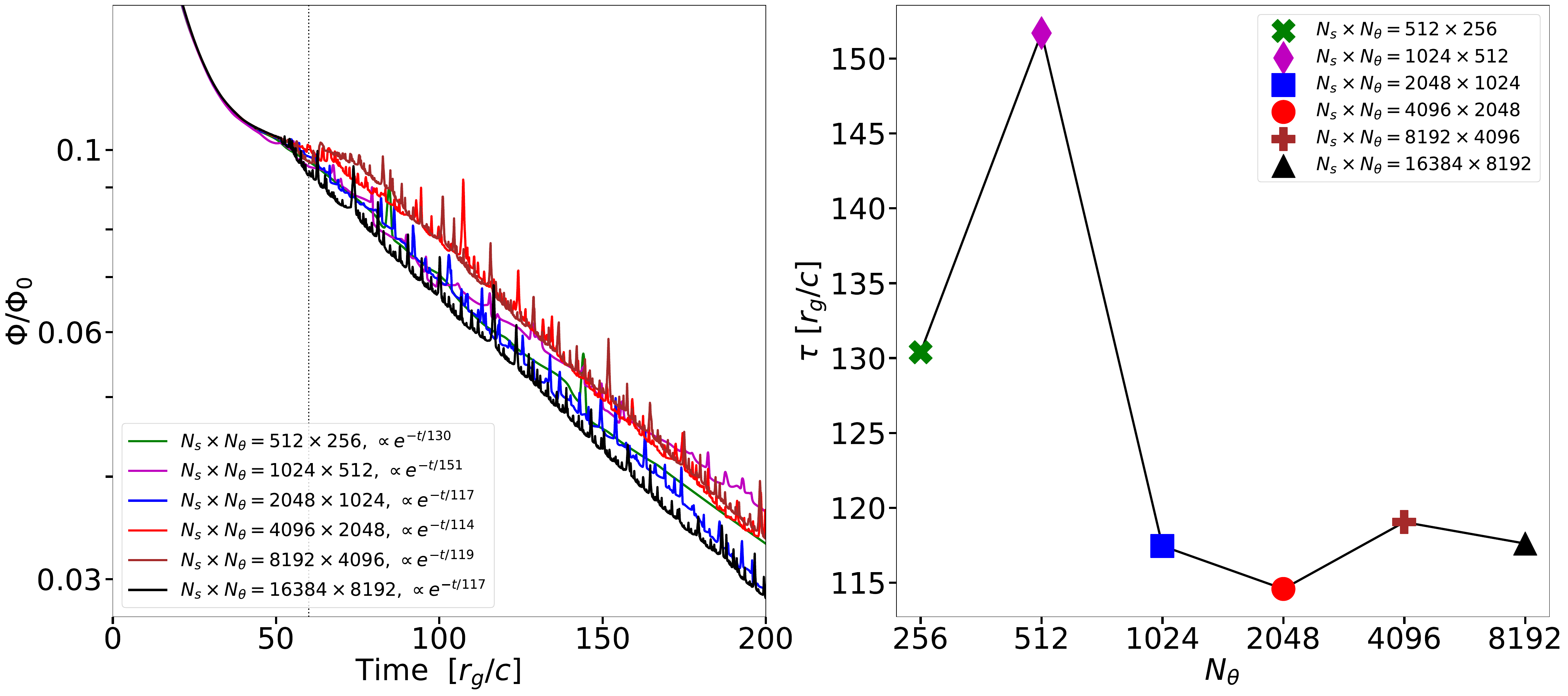}
}
\caption{   
Magnetic flux evolution of six 2D simulations of an non-spinning BH with an initial dipole magnetic field geometry with increasingly higher resolutions showing numerical convergence. The left panel displays the time evolutions of the total instantaneous event horizon absolute magnetic field flux $\gls{magnflux}_{\textrm{tot}}(\gls{time})$ normalized by the initial total event horizon absolute flux $\gls{magnflux}_{0}= \gls{magnflux}_{\textrm{tot}} (\gls{time}=0)$. 
The vertical dotted line in both panels shows the time $60 \ \gls{gravradius} / \gls{spol}$ at which pressure equilibrium is established for all simulations, thereby separating the first pressure equilibration phase from the second pressure balance phase. 
The right panel displays the exponentially fitted flux decay timescales versus the resolution shown by the number of computational cells in the direction polar. 
} 
\label{fig:numconv}
\end{figure}

To investigate the numerical convergence of our simulations we performed six 2D simulations of a non-spinning BH with an initial dipole magnetic field geometry for six increasingly higher numerical resolutions. 
We use logarithmic Kerr-Schild coordinates for the computational grid with a uniform base resolution of $\gls{num}_{\gls{mksradius}} = 512$ and $\gls{num}_{\gls{sphpolar}} = 256$ computational cells along the radial $\gls{mksradius}$ and polar $\gls{sphpolar}$ coordinates respectively and increase resolution by doubling the number of cells in each coordinate direction. 
We base our convergence on the evolution of the total instantaneous absolute magnetic field flux $\gls{magnflux}_{\textrm{tot}}(\gls{time}) = \int_{0}^{2 \pi} \int_{0}^{\pi} |\gls{Bcomp}^{\gls{mksradius}}(\gls{time})|\sqrt{\gls{spmetricdet}} d\gls{sphpolar}d\gls{sphazi}$ over the event horizon normalized by the initial total event horizon absolute flux $\gls{magnflux}_{0}= \gls{magnflux}_{\textrm{tot}} (\gls{time}=0)$  (see in Figure \ref{fig:numconv} in the right panel the colored continuous lines). 
In the first phase for $\gls{time} \lesssim 60 \ \gls{gravradius} / \gls{spol}$ a large drop in flux occurs which is dominated by poloidal field loops falling into the BH. 
This process occurs on global scales such that all simulated resolutions show equivalent flux evolution. 
During this first phase the initial dipole field geometry transitions into a split monopole geometry. 
Once the split monopole has formed the second phase starts in which the flux decays exponentially through magnetic reconnection in the equatorial current sheet. 
For a resolution of $\gls{num}_{\gls{mksradius}} \times \gls{num}_{\gls{sphpolar}} = 2048 \times 1024$ cells or higher, the variability in the flux evolution increases dramatically. This can be attributed to the abundant formation of plasmoids containing magnetic flux and shows up as a momentary increase of magnetic flux when plasmoids are passing over the event horizon. 

In the plasmoid-mediated reconnection regime the reconnection rate is converged to the asymptotic rate. Therefore, although in our ideal MHD simulations magnetic reconnection occurs at the grid-scale the magnetic energy dissipation and flux decay are converged, for sufficiently high resolutions.
The flux decay is fitted using an exponential function $\propto e^{-\gls{time}/\tau}$ where $\tau$ is the flux decay timescale (see in Figure \ref{fig:numconv} in the left panel the colored dashed lines). 
Figure \ref{fig:numconv} shows the decay timescales versus the resolution in the right panel. 
This shows that for a resolution $\gls{num}_{\gls{mksradius}} \times \gls{num}_{\gls{sphpolar}} = 2048 \times 1024$ cells or higher, the flux decay time scale is converged to its asymptotic value. 
The slight variation in decay timescales for resolutions of $\gls{num}_{\gls{mksradius}} \times \gls{num}_{\gls{sphpolar}} = 2048 \times 1024$ cells or higher can likely be attributed to the variations of the chaotic plasmoid dynamics between initially equivalent simulations. 
Therefore, we conclude that our simulations converge on asymptotic reconnection rate and flux decay for a resolution of $\gls{num}_{\gls{mksradius}} \times \gls{num}_{\gls{sphpolar}} = 2048 \times 1024$ cells.

\end{document}